FRONT MATTER
**Title**
**Nanoscale ferroelastic twins formed in strained LaCoO$_3$ films**

**Authors**
Er-Jia Guo,[1, 2, 3, *] Ryan Desautels,[1] David Keavney,[4] Manuel A. Roldan,[5] Brian J. Kirby,[6] Dongkyu Lee,[1] Zhaoliang Liao,[1] Timothy Charlton,[1] Andreas Herklotz,[1, 7] T. Zac Ward,[1] Michael R. Fitzsimmons,[1, 8] and Ho Nyung Lee [1, *]

**Affiliations**
[1]Oak Ridge National Laboratory, Oak Ridge, TN 37831, United States
[2]Beijing National Laboratory for Condensed Matter Physics and Institute of Physics, Chinese Academy of Sciences, Beijing 100190, China
[3]Center of Materials Science and Optoelectronics Engineering, University of Chinese Academy of Sciences, Beijing 100049, China
[4]Advanced Photon Source, Argonne National Laboratory, Argonne, Illinois 60439, United States
[5]Eyring Materials Center, Arizona State University, Tempe, AZ 85287, United States
[6]NIST Center for Neutron Research, National Institute of Standards and Technology, Gaithersburg, Maryland 20899, United States
[7]Institute for Physics, Martin-Luther-University Halle-Wittenberg, Halle (Saale) 06120, Germany
[8]Department of Physics and Astronomy, University of Tennessee, Knoxville, TN 37996, United States

Corresponding authors: hnlee@ornl.gov and ejguo@iphy.ac.cn

**Abstract**.
The coexistence and coupling of ferroelasticity and magnetic ordering in a single material offers a great opportunity to realize novel devices with multiple tuning knobs. Complex oxides are a particularly promising class of materials to find multiferroic interactions as they often possess rich phase diagrams and the interactions are very sensitive to external perturbations. Still, there are very few examples of these systems. Here we report the observation of twinning domains in ferroelastic LaCoO$_3$ epitaxial thin films and their geometric control of structural symmetry that are intimately linked to the material's electronic and magnetic states. A unidirectional structural modulation is achieved by selective choice of substrates possessing two-fold rotational symmetry. This modulation perturbs the crystal field splitting energy, leading to unexpected in-plane anisotropy of orbital configuration and magnetization. These findings demonstrate the utilization of structural modulation to control multiferroic interactions and may enable a great potential for stimulation of exotic phenomena through artificial domain engineering.

**MAIN TEXT**
**Introduction**

Ferroelastics are the largest class of ferroic materials and are essential for many applications, such as vibration sensors, smart mechanical switches, and acoustic devices (*1-3*). Complex perovskite oxides with a rhombohedral lattice structure offer a particularly interesting subclass of ferroelastic materials. In this structure, the ferroelastic response is driven by a distortion of the cubic parent structure through stretching along one of the four body diagonals of the perovskite unit cell. To minimize the total elastic energy, ferroelastic oxides commonly form twining domains at the expense of interfacial energy associated with domain walls (DWs) (*4-5*). The orientation of these DWs is strictly constrained by a highly directional bonding in the crystalline matrix. Arrays of periodic domains have been reported in ferroelastic rare-earth phosphates (*1, 3, 6-8*) and aluminates (*3, 5, 9, 10*). The ferroelastic domains walls passing through the single crystals are invariably arranged regularly due to the strain compatibility conditions between the adjacent domains, forming a quasi-one-dimensional domain configuration. The unique periodic ferroelastic domains have only been observed in the bulk previously. In addition, conventional ferroelastic materials also typically lack ferromagnetic ordering (*1, 3, 5, 6-10*), which precludes direct coupling between the ferroelastic and ferromagnetic order parameters.

LaCoO$_3$ (LCO) is a ferroelastic perovskite oxide. Interestingly, unlike the bulk LCO, the epitaxially strained LCO thin films (*11, 12*) exhibit the emergent ferromagnetism at low temperatures, attracting increasing attention recently (*13-19*). Thus, ferroelastic LCO films provide an ideal platform to investigate the coupling between ferroelasticity and magnetism. The delicate interplay between the crystal field splitting energy ($\Delta_{cf}$) and inter-atomic exchange interaction energy ($\Delta_{ex}$) determines the active spin crossover between low and high spin states of Co ions. Since $\Delta_{cf}$ is extremely sensitive to the changes of the Co-O bond length and the Co-O-Co bonding angle (*18, 19*), a small structural perturbation by strain can significantly modify the Co spin state, thereby affecting the magnetism of LCO films. Misfit strain between an epitaxial ferroelastic thin film and a high-order-symmetry substrate is accommodated in different ways. Typically, biaxial strain relaxation originated from the lattice mismatch proceeds through elastic deformation of the unit cell or the formation of misfit dislocations at the film/substrate interface. However, misfit shear strain relaxes via spontaneous symmetry reduction of a film upon the ferroelastic transition, favoring the formation of ferroelastic domains. (*11*) In the thin films, the orientation of ferroelastic domains is strongly influenced by the morphology of underlying substrates. For vicinal substrates, the miscut direction ($\alpha$) and miscut angle ($\beta$) are two essential parameters that determine the in-plane domain configuration ([Fig. 1a](#)). For a large $\alpha$, the four-fold symmetry of the (001) plane is broken into two-fold by exposing both (010) and (100) facets at the terrace steps. Therefore, the ferroelastic thin films have two possible in-plane epitaxial directions, resulting in multiple domain formation with random orientations. However, if the $\alpha$ is close to zero, ferroelastic thin films grow along a preferred direction [either (100) or (010)], thus the unidirectional structural twinning arrangement can be realized. Such an approach allows us to investigate the effect of a single structural modulation on the intriguing physical properties of ferroelastic thin films.

Here, using LCO as an example, we show the stabilization of one-dimensional (1D) and checker-board-like twinning domains in ferroelastic thin films by utilizing the morphology and symmetry of underlying substrates. We demonstrate for the first time that a small change of the surface miscut direction or the choice of crystallographic symmetry play a crucial role in the formation of ferroelastic twinning domains. The structural modulation induces a large anisotropy in the orbital occupancy, accompanied by an emergence of robust in-plane magnetic anisotropy. Our work strengthens the understanding of strong correlation between the ferroic order parameters in a multiferroic.

**Results and discussion**

LCO thin films with a thickness of 35-unit cells (u.c.) were grown on (001)-oriented $TiO_2$-terminated $SrTiO_3$ (STO), $(LaAlO_3)_{0.3}(SrAl_{0.5}Ta_{0.5}O_3)_{0.7}$ (LSAT), and (110)-oriented $NdGaO_3$ (NGO) substrates by pulsed laser deposition (see Methods). The LCO films were subsequently capped with an ultrathin STO layer with a thickness of 5-u.c. to prevent the formation of nonstoichiometric surface due to oxygen vacancies (*17, 20*). Results from X-ray diffraction (XRD) measurements confirm an excellent epitaxial growth of all layers. Reciprocal space maps (RSMs) were measured around the film's 103 reflection (pseudocubic notation is used throughout the article) of an LCO film through azimuthally rotating the sample by 90° with respect to the surface's normal (Fig. 1c). RSM results confirm that the LCO film is coherently grown on the STO substrate. As the lattice parameter of STO is larger than that of bulk LCO, the LCO film is under biaxial tensile strain. The peak position of the 013 ($0\bar{1}3$) reciprocal lattice reflection of LCO film is shifted downward (upward) along the *l*-direction with respect to the 103 and $\bar{1}03$ reflections. This result indicates that the symmetry of LCO film grown on a cubic substrate changes from the bulk rhombohedral (*R*3*c*) to a symmetric monoclinically distorted lattice structure (*I*2/*a*). The reflections of LCO film show a central peak and two symmetric satellite peaks, which originate from ferroelastic twinning domains. We find the two satellite peaks only appear in the RSMs for the 103 and $\bar{1}03$ reflections, whereas the 013 and $0\bar{1}3$ reflections do not have these satellite peaks. Rocking curve scans around LCO 002 reflection were recorded as a function of the in-plane rotation angle ($\varphi$) (Fig. 1d). A cosine-like modulation of the individual satellite peak position is observed. This observation confirms that the structural modulation is unidirectional. Furthermore, we only observe the first-order satellite peak of each 00*l* LCO reflection, indicating a rather short correlation length of twinning domains. The angular difference ($\omega$-$\omega_{00l}$) between the central peaks and satellite peaks linearly reduces as the diffraction order (*l*) increases, revealing a constant in-plane structural modulation (*15*). We also measured reciprocal space maps (RSMs) around different $h03$ ($h$ = 0, 1, 2, 3) reflections of the STO substrate (Extended data Fig. S1). The peak splitting of the LCO film along the $h$-direction is the same for all RSMs; however, no peak splitting occurs along the $l$-direction. This observation further reinforces the presence of a periodic structural modulation along the in-plane direction only (Fig. 1a). From the first order rocking curve scan, we can obtain the tilt angle $\gamma$ [= 2.2(1)°] between two twinning domains (Extended data Fig. S1). Moreover, by calculating the satellite spacing ($\Delta q_x$) between the central peaks and first-order satellite peaks, we can determine the periodicity ($\zeta = 1/\Delta q_x$) of twinning domains to be ~ 10 nm (Fig. 1b). Note that the XRD results were obtained over a large portion of the sample (~ tens of mm$^2$), indicating the 1D twining domains are prevalent throughout the LCO film.

To examine the correlation between the structural properties and surface morphology in our LCO films, the topography was measured over an area of 5×5 µm$^2$ by atomic force microscopy (AFM) at several places of a sample (Extended data Fig. S2). All images show clear step-and-terrace patterns with straight, continuous, and parallel to the [010] direction (*21*), *i. e.* $\alpha \approx 0$. We also find $\zeta$ increases as $\beta$ increases (Extended data Fig. S3). This is because the density of facet planes at steps increases with $\beta$, thereby the ferroelastic twinning domains are more easily stabilized on the vicinal substrates with a larger $\beta$ (*22, 23*). In addition, $\zeta$ is approximately proportional to $t_{LCO}^{1/2}$ (Extended data Fig. S4c), where $t_{LCO}$ is the film thickness, in accordance with thermodynamic consideration for domain formation in epitaxial ferroelastic films (*24, 25*). The substrate morphology—as defined by $\alpha$ and $\beta$—controls the formation and arrangement of the unidirectional twinning domains.

To explore the correlation between the electronic state and structural distortion in the LCO films, we performed element-specific X-ray absorption spectroscopy (XAS) in total fluorescence yield (FY) mode. The features at the Co $L_{3,2}$-edge confirm that our films are oxygen stoichiometric with mixed spin state Co$^{3+}$ ions (*17, 26, 27*). The contribution from Co$^{2+}$ ions induced by oxygen vacancies is negligible (< 2%) in our films (*20*). X-ray linear dichroism (XLD) was measured to compare the difference in the electronic occupancy of Co $d$ orbitals. To ensure that the observed linear dichroism is from the anisotropy in the $e_g$ band, the measurements were taken under zero magnetic field (*28*). XLD measurements were performed with the X-ray scattering planes parallel to the (100) and (010) planes, respectively (Fig. 2a). The XAS intensities from the out-of-plane ($E_{oop}$//[001]) and in-plane ($E_{ip}$//[100] or [010]) linearly polarized X-ray beam are proportional to the density of unoccupied states, *i. e.* holes, in the Co $d_{3z^2-r^2}$ and $d_{x^2-y^2}$ orbitals, respectively (*26-28*). As shown in Fig. 2b, the peak energy of $I_{ip}$ is lower than that of $I_{oop}$, thus the occupation of $d$ electrons in the $d_{x^2-y^2}$ orbital is larger compared to the $d_{3z^2-r^2}$ orbital. The difference in the peak positions between $I_{ip}$ and $I_{oop}$ implies a splitting of $e_g$ band arising from the $d_{x^2-y^2}$ and $d_{3z^2-r^2}$ orbitals (*28*). Direct inspection of the peak positions for XAS curves reveals that, for the (010) scattering geometry, the $I_{oop}$ is nearly the same as the $I_{ip}$. For the (100) scattering geometry, the different in the peak position between the $I_{oop}$ and $I_{ip}$ increases to ~ 0.15 eV, indicating a large $e_g$ band splitting. These observations imply that the electronic configuration exhibits a significant anisotropy.

The difference between $I_{oop}$ and $I_{ip}$ is shown in Fig. 2c. The peak intensity of XLD spectra for the (010) scattering geometry is remarkably different from that of the (100) scattering geometry. To obtain a quantitative estimate of the imbalance in $e_g$ band occupation, we applied the sum rule for linear dichroism to calculate the orbital polarization $P = (n_{x^2-y^2} - n_{3z^2-r^2})/(n_{x^2-y^2} + n_{3z^2-r^2})$, [where $n_{x^2-y^2}$ and $n_{3z^2-r^2}$ represent the numbers of electrons] (*29*). $P$ for the (010) scattering plane is ~ 27%, which is almost two times larger than that for the (100) scattering plane (~ 14%). The monoclinic distorted lattice will lead to a distortion of CoO$_6$ octahedral, *e. g.* changes of bonding angle and bond length, resulting in a strong anisotropy of electron occupancy along different planes. These observations are strong evidence that the unidirectional structural modulation induces the large anisotropy in the electronic configuration, affecting both band splitting and orbital polarization in the Co $d$ bands.

In addition, we collected the total electron yield (TEY) spectra (not shown), which are known to be surface sensitive with probing depth of ~ 3-5 nm. The results from the FY and TEY are identical, demonstrating that the strong anisotropy in the electronic states arises from the structural modulation of the entire LCO film and is not limited to the surface.

The unidirectional structural distortion has a strong influence on the magnetization of LCO films. As shown in Fig. 3a, the magnetization (*M*) exhibits square-like hysteresis loops, corroborating the ferromagnetic order in tensile-strained LCO films (*13-17*). Hence, *M*(*H*) does not saturate even for the fields of ± 7 T, suggesting an additional paramagnetic (PM) contribution. Previously, the PM component has been observed in LCO thin films and single crystals (*13, 30*), where two magnetic sublattices contribute to the total magnetization. In-plane *M*(*H*) loops show clear anisotropy with a larger coercive field ($H_C$) and a higher *M* for *H* // [010] compared to those parameters for *H* // [100]. The inset of Fig. 3b shows the in-plane angular dependence of *M* at 10 and 70 K. A strong sine modulation of *M* is observed at 10 K with a maximum and minimum *M* along the [010] and [100] directions, respectively. The change of *M* is strongly correlated with the structural modulation. However, the variation in *M* is subtle at 70 K. *M*(*T*) curves were measured while warming the sample with *H* // [010], [100], and [001], as shown in Fig. 3b. For *H* // [010], *M*(*T*) exhibits a sharp transition at the Curie temperature ($T_C$) of ~ 75 K (*13-17*). Surprisingly, *M*(*T*) shows two distinct magnetic transitions when *H* // [100]: One at $T_C$ and another at around 60 K. We note that the trends of *M*(*T*) curves are quite similar to each other for *H* // [100] and [001] when *T* < 60 K.

The unique magnetic behavior in our LCO films could be attributed to the unconventional electronic states triggered by the unidirectional structural modification. Since $\Delta_{cf} \neq 0$ for tensile-strained LCO films, the Co–O molecular orbitals are split into three-fold degenerate $t_{2g}$ and two-fold degenerate $e_g$ bands (Figs. 3c–3e). The electronic configuration, *i.e.* the spin state of Co ions, is controlled by the energy difference $\Delta E = \Delta_{cf} - \Delta_{ex} - W/2$, where *W* is the bandwidth (*W*) between the hybridized Co $e_g$ orbital and O 2*p* orbital. In unidirectionally distorted $CoO_6$ octahedra, the bond length ($d$) and bonding angle ($\delta$) exhibit strong anisotropy, thus both $W$ $[\propto \frac{\cos(\pi-\delta)}{d^{3.5}}]$ and $\Delta_{cf}$ ($\propto \frac{1}{d^5}$) depend on the crystallographic orientation. We thus expect the spin state of Co ions to depend upon crystallographic orientation. The effective paramagnetic moments ($\mu_{eff}$) ~ 4.67(3) and ~ 4.12(2) $\mu_B$/Co (where $\mu_B$ is the Bohr magneton) were obtained from the susceptibilities above 100 K along the [010] and [100] orientations (Fig. 3b), respectively. Assuming a single-electron model, $\mu_{eff}$ is equal to $g_e \times \sqrt{S(S+1)}$ $\mu_B$/Co, where the electron $g$ factor $g_e = 2$. We calculate the corresponding effective spin state of $Co^{3+}$ ion is $S_{[010]} = 1.89(3)$ $\mu_B$/Co and $S_{[100]} = 1.61(2)$ $\mu_B$/Co. The large value of $S_{[010]}$ and $S_{[100]}$ can only be achieved with occupancy > 89% and > 61%, respectively, of high spin (HS, $S = 2$) Co ions. In addition, the XLD results demonstrate the $e_g$ band splitting along the [100] orientation is ~ 0.15 eV larger than that along the [010] orientation. Thus, electrons preferentially occupy the lower energy orbitals, resulting in a higher spin state along the [010] orientation compared to that of [100] orientation. These observations are consistent with the magnetization measurements.

To further illustrate the importance of the two-fold rotational symmetry imposed by the substrates, we have grown LCO films on (110)-oriented orthorhombic NGO substrates (*19, 34*). The in-plane lattice constants along the [$\bar{1}$10] and [001] orientations are different (Fig. 4a), providing intrinsic anisotropic misfit strain. The structural anisotropy leads to an asymmetric LCO lattice structure, similar to the LCO films grown on step-and-terrace STO substrates. XRD measurements confirm the formation of 1D twinning domains in LCO films along the [$\bar{1}$10] orientation of NGO (Extended data Fig. S5). Therefore, the unidirectional structural distortion in LCO films can be achieved either by choice of a substrate morphology, *i.e.,* the miscut direction (α) and miscut angle (β), or the choice of substrate crystallographic symmetry. Due to the strong PM background from NGO and the large coercive field of LCO films, the magnetic properties of LCO films cannot be easily quantified with magnetometry. The magnetization, however, can be elucidated by X-ray magnetic circular dichroism (XMCD) (*35*) and quantified with polarized neutron reflectometry (PNR) (*36*). Fig. 4b shows XMCD spectra of the Co *L*-edge with a magnetic field of 5 T applied along the [$\bar{1}$10] and [001] directions at 10 K. The difference in XMCD signals provides solid evidence for the in-plane anisotropy. A direct comparison of absolute magnetic moment along the two in-plane directions was obtained from PNR measurements at 10 K with a magnetic field of 3 T (Extended data Fig. S6). The nuclear (atomic density) and magnetization depth profiles of the LCO heterostructure are shown in Fig. 4c and 4d, respectively. The magnetic moment of LCO film along the [001] orientation [0.70(5) $\mu_B$/Co] is larger than that along the [$\bar{1}$10] orientation [0.54(2) $\mu_B$/Co]. Both techniques reveal an appreciable difference in the magnetic responses along different in-plane directions, which offers further confirmation that the magnetic ground states are strongly modulated by the orientation of unidirectional twinning domains. Thus, choice of substrate symmetry and modification of surface morphology, which controls the periodicity and orientation of twinning domains, is an effective means to control the magnetic properties of LCO films.

Since we have found the 1D twinning domain formation originates from the terraced surfaces, one can hypothesize that diminishing step terrace features on a substrate may alter the formation of 1D twinning domains. Indeed, we were able to stabilize checkerboard-like twinning domains in LCO films grown on a cubic LSAT substrate, which lacks well-defined step-and-terrace surface due to the cation segregations (Extended data Fig. S7). The formation of twinning domains along both [010] and [100] directions were observed, confirming the electronic states of the LCO films exhibit a strong in-plane anisotropy with a reduced electron occupancy of the $d_{3z^2-r^2}$ orbital along the (110) plane (Extended data Fig. S8). The magnetic response for fields along the [100] and [010] orientations were identical, whereas the largest magnetization was observed with field along the diagonal direction (Extended data Fig. S9). These observations further reinforce the strong correlation between structural distortion and electronic/magnetic ground states in strained LCO films.

Finally, we demonstrate that artificially designed step-and-terrace morphology with perfectly aligned steps along either [010] or [100] orientation is essential for the formation of nanoscale 1D twinning domains. A slight change in the miscut direction towards the diagonal direction modifies the area ratio between the (010) and (100) facet planes at the steps. If the substrate's surface is covered by the two mixed facet planes, then thin films will grow orthogonally oriented checker-board like twinning domains, yielding isotropic in-plane ferromagnetism (Extended data Fig. S10). The isotropic magnetization of these LCO films is determined by the average from the maximum and minimum magnetizations of LCO films of individual domains.

**Conclusions**

In summary, we report the formation of ferroelastic twinning domains in $LaCoO_3$ epitaxial thin films by utilizing the substrate morphology and crystallographic symmetry. Surface-modified cubic substrates or orthorhombic substrates provide opportunities to achieve fine control of unidirectional structural modification, which provides the fine control over the mangnetic anisotropy. The competition between minimizing the elastic strain energy at the expense of the domain wall energy due to the induced two-fold rotational symmetry leads to the formation of 1D twinning domains. Our results demonstrate direct transfer of unique symmetry-imposed domain pattern from the structure into the electronic state (orbital occupancy) and subsequent magnetic order. Utilizing the control of domain architectures to investigate the strong correlation between different ferroic orders opens an avenue towards an improved fundamental understanding of the strongly correlated interactions for future functional device applications.

**Materials and Methods**

**Thin film synthesis and physical property characterization.** LCO thin films were grown by pulsed laser deposition. Prior to thin film deposition, the SrTiO$_3$ (STO) (001) substrates and NdGaO$_3$ (NGO) (110) substrates were treated by buffered hydrofluoric acid and subsequent annealing at 1050 °C for 2 hours to achieve atomically flat surfaces. The (LaAlO$_3$)$_{0.3}$(SrAl$_{0.5}$Ta$_{0.5}$O$_3$)$_{0.7}$ (LSAT) substrates were untreated due to the cation segregation, thus, no specific termination and step-and-terrace are formed on LSAT substrates. During film growth, the substrate temperature was kept at 700 ºC, the oxygen partial pressure was maintained at 100 mTorr, and the laser fluence was fixed at ~ 1.2 J cm$^{-2}$. LCO films (~ 35 u.c.) grown on STO and LSAT were capped with an ultrathin STO layer (~ 5 u.c.) to prevent the formation of oxygen vacancies at the surface. LCO films (~ 80 u.c.) were grown on NGO and further capped with a STO layer (~ 80 u.c.) for the neutron reflectivity measurements. After growth, the samples were cooled to room temperature in 100 Torr of oxygen to ensure the right oxygen stoichiometry. X-ray reflectivity (XRR) measurements were performed to check the layer thickness and interface abruptness. The crystalline quality of all layers was checked by XRD *θ-2θ* and rocking curve scans. The in-plane strain states of the films were characterized by XRD reciprocal space mapping (RSM). The morphologies of substrates and samples were characterized with a Nanoscope III atomic form microscopy (AFM) in the tapping mode. The magnetic properties of LCO films were measured using a SQUID magnetometer (Quantum Design). The magnetization of the LCO films was acquired by subtracting the diamagnetic signals from the substrates (STO and LSAT).

**Soft X-ray absorption spectroscopy (XAS)**. XAS experiments were performed at the beamline 4-ID-C of the Advanced Photon Source at Argonne National Laboratory. The valence state of Co ions in the as-grown LCO films was directly probed by XAS measurements. XAS spectra near the Co *L*-edges were measured with both bulk-sensitive fluorescence yield (FY) mode and surface-sensitive total electron yield (TEY) mode at 10 K. The incident angle between the X-ray beam and sample's surface plane is around 30º. X-ray linear dichroism (XLD) was measured by linearly polarized X-rays with polarization vector ($E$) in parallel to the in-plane or out-of-plane direction of the sample, respectively. X-ray magnetic circular dichroism (XMCD) measurements were performed under an in-plane magnetic field of ± 5 T. The XMCD signals were calculated from the difference in the absorption of the right- and left-hand circularly polarized X-rays. The XMCD signal flips its polarity with reversal of applied magnetic field, indicating that the tensile-strained LCO films are magnetic in origin. The XMCD signals were corrected by multiplying 96%/cos(30º) ~ 1.1 with considering the incident angle (30º) and circular polarization (96%) of the X-ray beam. XLD and XMCD measurements were repeated by successively rotating the sample in 90º, e. g., along two perpendicular in-plane orientations of the films.

**Polarized Neutron Reflectometry (PNR).** PNR experiments on LCO films grown on NGO substrates were performed at the PBR beamline of the NIST Center for Neutron Research (NCNR). The samples were field-cooled and measured with magnetic fields of 0.7 (results were not shown) and 3 T, respectively. The magnetic fields were applied along the in-plane directions, e.g. [001] and [$\bar{1}$10], of the NGO substrate. PNR measurements were conducted at 10 K in the specular reflection geometry with the wave vector transfer ($q$) perpendicular to the sample's surface plane. We measured the spin-up ($R^+$) and spin-down ($R^-$) neutron reflectivity as a function of $q$ (=$4\pi sin\theta_i/\lambda$), where $\theta_i$ is the incident angle of neutrons and $\lambda$ is the neutron wavelength. The neutron reflectivity was normalized to the asymptotic value of the Fresnel reflectivity $R_F$ (= $16\pi^2/q^4$) in order to better illustrate the small divergence between two measuring geometries. To separate the nuclear information from the magnetic scattering, the data were also present as the spin-asymmetry SA (= [($R^+ - R^-$)/($R^+ + R^-$)]). The raw data were shown in the Supporting information. Data fitting was performed using both GenX and NIST Refl1D program (*37*). The results from two fitting programs are in good agreement. We constrained our chemical depth profile from fitting a model to XRR data to deduce the magnetization depth profile of the samples. The in-plane magnetization of the LCO films along the [001] and [$\bar{1}$10] orientations can be quantitatively determined.

**Supplementary Materials**

**Fig. S1**. Structural properties of a LCO film.

**Fig. S2**. Topography of a LCO film capped with a STO ultrathin layer.

**Fig. S3**. Evolution of the 1D twinning domain in LCO films grown on STO substrates with different miscut angles.

**Fig. S4**. Thickness dependent twinning domain periodicity in LCO films.

**Fig. S5**. 1D twinning domain in LCO films grown on NGO substrates.

**Fig. S6**. In-plane magnetic anisotropy in LCO films grown on NGO substrate.

**Fig. S7**. Checkerboard-like twinning domains observed in LCO films on LSAT substrates.

**Fig. S8**. Magnetic properties of LCO film on LSAT substrate with checkerboard-like twinning domains.

**Fig. S9**. Anisotropic electronic states in LCO films on LSAT substrates.

**Fig. S10**. Comparison of the in-plane magnetic anisotropy in LCO films with and without (w.o.) 1D twinning domains.

**Acknowledgements**

We thank Yaohua Liu, Feng Ye, Changhee Sohn, Hyoungjeen Jeen, Jong K. Keum, and Fernando A. Reboredo for valuable discussions. **Funding**: This work was supported by the U.S. Department of Energy (DOE), Office of Science, Basic Energy Sciences (BES), Materials Sciences and Engineering Division. The PNR measurements at the National Institute of Standards and Technology (NIST) Center for Neutron Research (NCNR), U.S. Department of Commerce, was performed via a user proposal. This research used resources of the Advanced Photon Source, a U.S. DOE Office of Science User Facility, operated for the DOE Office of Science by Argonne National Laboratory under Contract No. DE-AC02-06CH1135 (XAS). During the manuscript revision, E.J.G. was funded through the Hundred Talent Program of Chinese Academy of Sciences. **Author contributions:** H.N.L. directed the project. E.J.G. and D.L. grew thin film samples, and E.J.G. carried out the structural and magnetic property measurements. E.J.G., R.D., and D.K. performed the XAS experiments. E.J.G. and B.J.K conducted the PNR measurements. D.L., M.A.R., Z.L., A.H., T.C., T.Z.W., M.R.F., H.N.L. contributed to the data analysis and interpretation. E.J.G., M.R.F., and H.N.L. wrote the manuscript with contributions from all the coauthors. **Competing interests:** The authors declare that they have no competing financial interests. **Data and materials availability:** All data needed to evaluate the conclusions in the paper are present in the paper and/or the Supplementary Materials. Additional data available from authors upon request. **Additional information:** Supplementary information is available in the online version of the paper. Reprints and permissions information is available online. Certain commercial equipment is identified in this paper to foster understanding. Such identification does not imply recommendation or endorsement by NIST, nor does it imply that the materials or equipment identified are necessarily the best available for the purpose.

**Figures and figure captions**

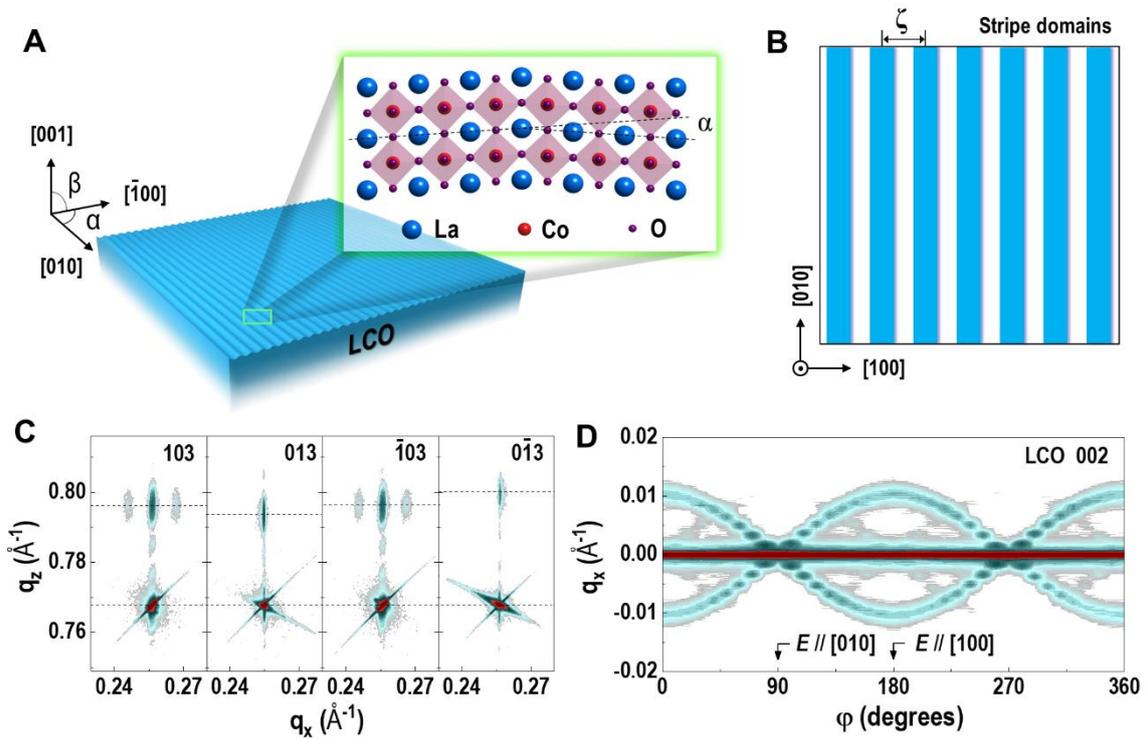

**Fig. 1 | 1D ferroelastic twinning domain is observed in a LaCoO$_3$ thin film**. (**A**). Schematic of 1D periodic twinning domains. Twinning domains constitute a spatially unidirectional structural modulation along the [100] orientation. The domains are parallel to the step edge direction of substrate. α and β are the miscut direction and miscut angle for a vicinal substrate, respectively. The inset shows a sketch of the monoclinically distorted LCO lattices at the ferroelastic domain wall. Notably, the tilt angle between two twinning domains is γ = 2.2 ± 0.1° derived from XRD measurements (Extended data Fig. S1). (**B**). Top-view of the stripe-like ferroelastic twinning domains. Two different colors represent differently oriented ferroelastic domains with an average periodicity ζ. (**C**). Reciprocal space maps (RSM) of a LCO film around the substrate's 103 reflection. RSMs are recorded by azimuthally rotating the sample with a step size of 90° with respect to the surface's normal. The LCO films have a monoclinically distorted lattice structure evidenced by the different $q_z$ spacing between the film's peak and substrate's peak. Two splitting satellite reflections at the same $q_z$ are shown for $h$03 reflections but are absent for 0$k$3 reflections. (**D**). Rocking curve scans around the LCO 002 reflections as a function of the in-plane rotation angle $\varphi$ in a step of 10°. The real space reflection angles are transformed into the reciprocal space wavevectors, from which we calculated ζ = 1/Δ$q_x$ ~ 10 nm. A cosine modulation of the satellite peak position indicates that the domain structure is stripy aligned perpendicular to the [100] orientation.

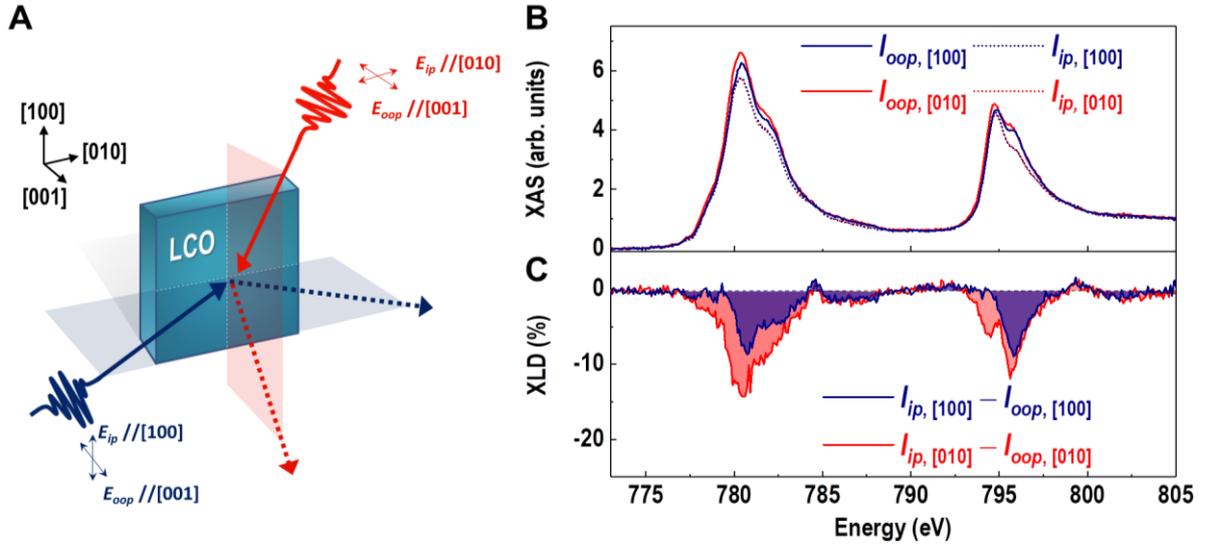

**Fig. 2 | Orbital polarization modulated by unidirectional structural distortions**.

(**A**). Schematic of the scattering geometry for XAS and XLD measurements with the X-ray beam aligned parallel to the (100) and (010) scattering planes. (**B**). XAS of a LCO film for the Co *L*-edge measured by the out-of-plane ($I_{oop}$, solid lines, $E_{oop}$//[001]) and in-plane ($I_{ip}$, dashed lines, $E_{ip}$//[100] or [010]) linearly polarized X-ray beams. (**C**). XLD of a LCO film for the Co *L*-edge. The XLD spectra indicate the hole occupancy in the $d_{3z^2-r^2}$ orbital is larger than that of the $d_{x^2-y^2}$ orbital for both measuring configurations. The degree of orbital polarization in the (010) plane is about two times larger than that in the (100) plane, indicating clear anisotropic orbital occupancy induced by 1D twinning domains. All spectra are collected and repeated more than four times with bulk-sensitive fluorescence yield (FY) detection mode at 10 K.

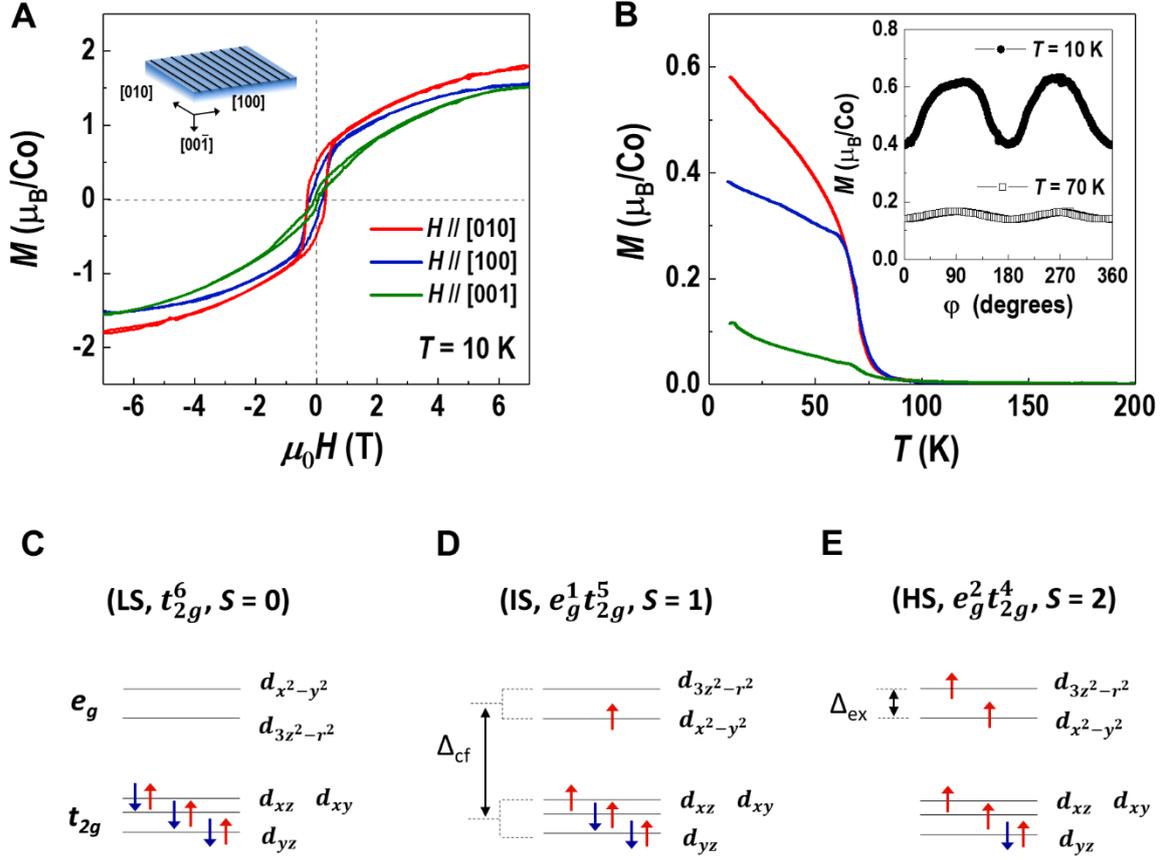

**Fig. 3 | In-plane magnetic anisotropy.** (**A**). *M-H* hysteresis loops and (**B**). *M-T* curves measured from a LCO film with *H* was applied along the [010], [100], and [001] directions. For the *M-T* curves, the cooling field was set at 0.1 T, and the measurements were carried out during the sample warm-up under a magnetic field of 0.1 T. The inset of (**B**) shows the angular dependence of the in-plane magnetization at 10 and 70 K under a magnetic field of 1 kOe. φ is the in-plane rotation angle with φ = 0° is *H* // [100] and φ = 90° is *H* // [010]. The in-plane magnetization is strongly modulated by the twinning domain arrangement. (**C**)–(**E**). Schematic energy-level diagrams of $Co^{3+}$ low-spin (LS), intermediate-spin (IS), and high-spin (HS) state configurations, respectively.

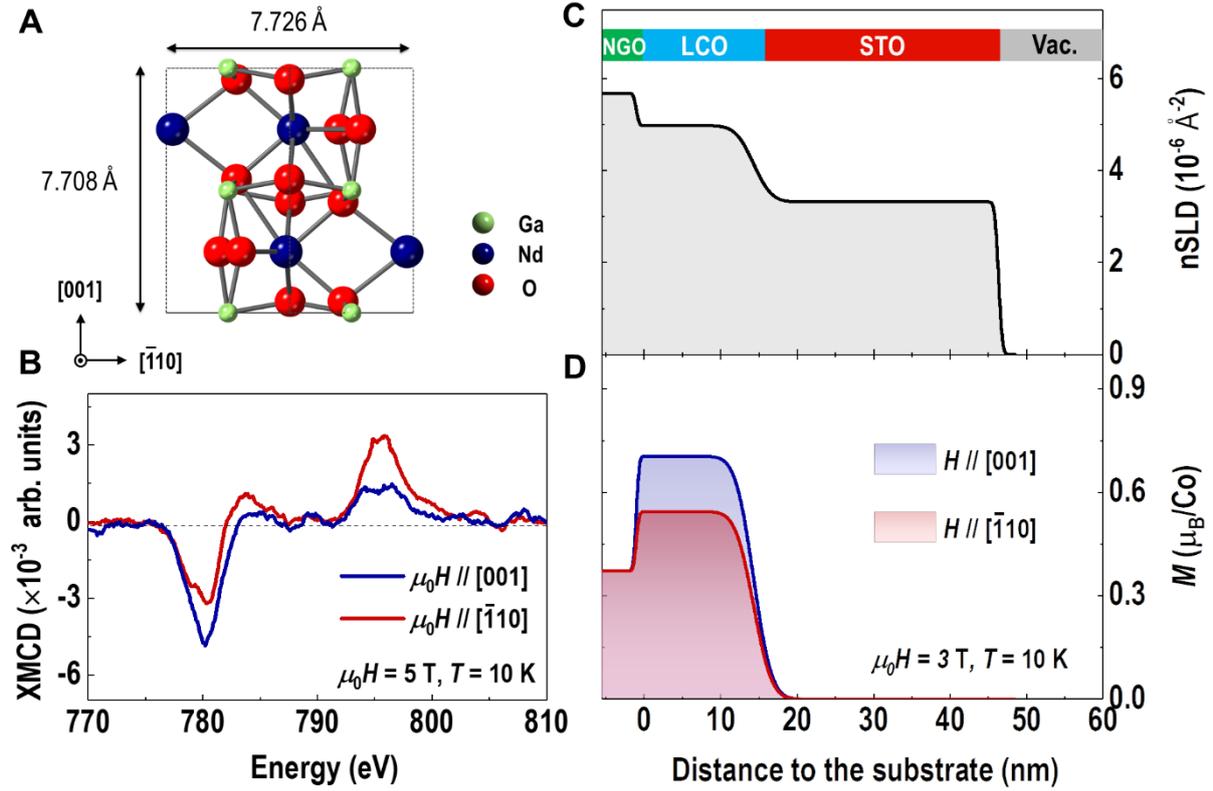

**Fig. 4 | In-plane magnetic anisotropy enforced by the orthorhombic substrate.** (**A**). The top-view lattice structure of orthorhombic (110)-oriented NdGaO$_3$(NGO). The in-plane lattice parameter along the [$\bar{1}$10] orientation is larger than that along the [001] orientation, leading to in-plane anisotropic shear strain in LCO films. (**B**). XMCD spectra for Co $L$-edge at 10 K with a magnetic field of 5 T applied along the [$\bar{1}$10] and [001] orientations. The XMCD signals were calculated from the difference between the $\mu^+$ and $\mu^-$ divided by their sum, as described by $(\mu^+-\mu^-)/(\mu^++\mu^-)$, where $\mu^+$ and $\mu^-$ denote XAS obtained from the right- and left-hand circular polarized photons, respectively. (**C**). Nuclear scattering length density ($n$SLD) and (**D**) magnetic moment (derived from the magnetic scattering length density, $m$SLD) depth profiles of a LCO film. The inset of (**C**) shows the schematic of the sample geometry. The LCO film was grown on a NGO substrate, then capped with a STO thin layer to prevent loss of oxygen at the LCO surface. PNR measurements were performed at 10 K after field cooling in 3 T. The magnetic field was applied along the [$\bar{1}$10] and [001] orientations, respectively.

**Supplementary Materials**

**Title**

**Nanoscale ferroelastic twins formed in strained LaCoO$_3$ films**

**Authors**
Er-Jia Guo,[1,2,3,*] Ryan Desautels,[1] David Keavney,[4] Manuel A. Roldan,[5] Brian J. Kirby,[6] Dongkyu Lee,[1] Zhaoliang Liao,[1] Timothy Charlton,[1] Andreas Herklotz,[1,7] T. Zac Ward,[1] Michael R. Fitzsimmons,[1,8] and Ho Nyung Lee [1,*]

**Affiliations**
[1]Oak Ridge National Laboratory, Oak Ridge, TN 37831, United States
[2]Beijing National Laboratory for Condensed Matter Physics and Institute of Physics, Chinese Academy of Sciences, Beijing 100190, China
[3]Center of Materials Science and Optoelectronics Engineering, University of Chinese Academy of Sciences, Beijing 100049, China
[4]Advanced Photon Source, Argonne National Laboratory, Argonne, Illinois 60439, United States
[5]Eyring Materials Center, Arizona State University, Tempe, AZ 85287, United States
[6]NIST Center for Neutron Research, National Institute of Standards and Technology, Gaithersburg, Maryland 20899, United States
[7]Institute for Physics, Martin-Luther-University Halle-Wittenberg, Halle (Saale) 06120, Germany
[8]Department of Physics and Astronomy, University of Tennessee, Knoxville, TN 37996, United States

Corresponding authors: hnlee@ornl.gov and ejguo@iphy.ac.cn

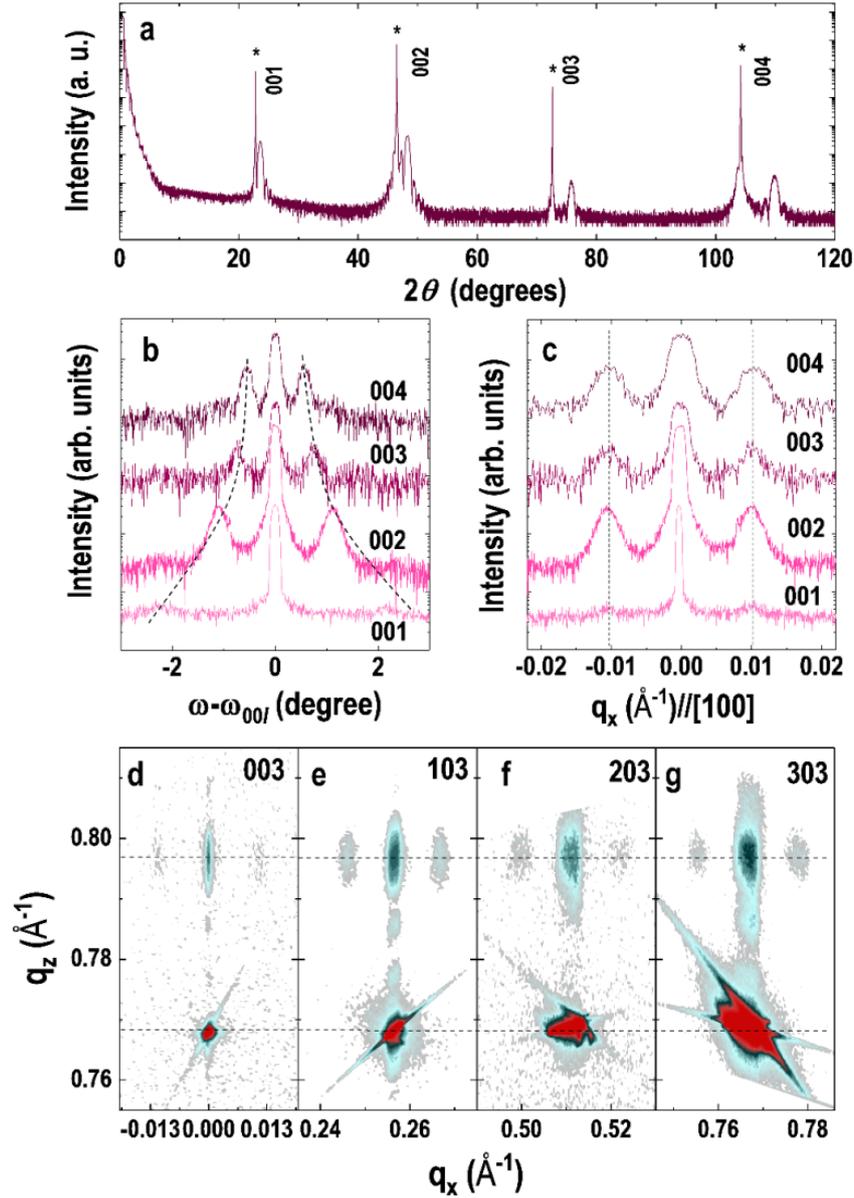

**Fig. S1 | Structural properties of a LCO film. a.** X-ray diffraction (XRD) $\theta$–$2\theta$ scan of a LCO film grown on a STO substrate (indicated with *). Distinct Laue thickness fringes around the (00$l$) Bragg peaks are observed, revealing the LCO film is of high structural quality. X-ray reflectivity (XRR) measurement confirms the uniform chemical composition within the LCO film and sharp interfaces (not shown). The thickness of the LCO film is ~ 13.1 nm (35 u.c.), in agreement with the thickness derived from the STEM analysis. **b.** Rocking curves around LCO 00$l$ ($l$ = 1, 2, 3, and 4) reflections. The rocking curves show intense zero-order peaks at $\omega_{00l}$ along with two symmetric splitting satellite peaks. The satellite peaks exhibit larger ($\omega$-$\omega_{00l}$) splitting values as the diffraction order ($l$) increases. This agrees with the behavior expected for the Bragg reflections arising from the tilted twinning domains. From the first order rocking curve, we could obtain the tilted angle ($\gamma$)

between the twinning domains is 2.2 ± 0.1°. **c**. Transformed rocking curves in reciprocal space lattice. All satellite peaks are at the same $q_z$ with $\Delta q_x \sim \pm 0.01$ Å$^{-1}$, corresponding to the twinning domain periodicity $\zeta$ (= $1/\Delta q_x$) of ~ 10 nm. **d – g.** RSMs around $h03$ ($h$ = 0, 1, 2, and 3) reflections. The splitting along the $h$-direction is a constant for all RSMs, while no splitting along $l$-direction, revealing the twinning domains are tilted along the in-plane direction only, not along the out-of-plane direction.

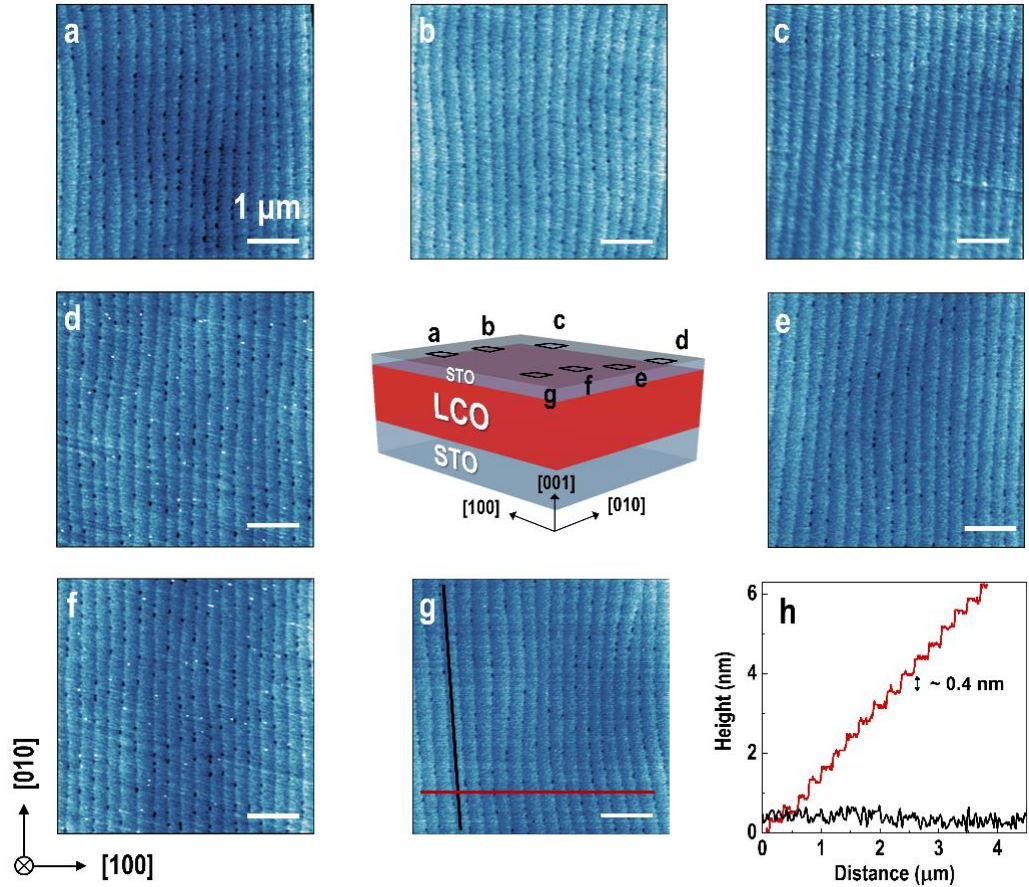

**Fig. S2 | Topography of a LCO film capped with a STO ultrathin layer.** Atomic force microscopy (AFM) images were recorded over an area of 5×5 µm$^2$ using a Veeco Dimension 3100 AFM operated in the tapping mode. Scale bars in **a** – **g** are 1 µm. The measurements were repeated at different locations, as marked in the sample schematic. The step-and-terrace feature was observed in all AFM images with an averaged r. m. s ~ 4.3 ± 0.2 Å, revealing a smooth sample surface over a large area. Line profiles along the black and red paths in **g** are plotted in **h**. The step height is about one pseudocubic unit cell (~ 0.4 nm) and the step width is about 250 ± 30 nm. The terrace is smooth with a root mean square roughness less than one pseudocubic unit cell. Notably, all steps are oriented along the [010] direction and perpendicular to the [100] direction. The terrace direction is strictly parallel to the orientation of the twinning domains. This result suggests the morphology of the artificially controlled steps would break the fourfold rotational symmetry in the (001) plane into twofold by exposing the (100) facet plane. Thus, the step-flow growth of LCO film preferentially in one direction, *e. g*., [100] orientation, in the present work. The monoclinically distortion of the LCO lattice structure will be strictly constrained to the growth direction of step flow, eventually forming the one-dimensional (1D) ferroelastic twinning domains.

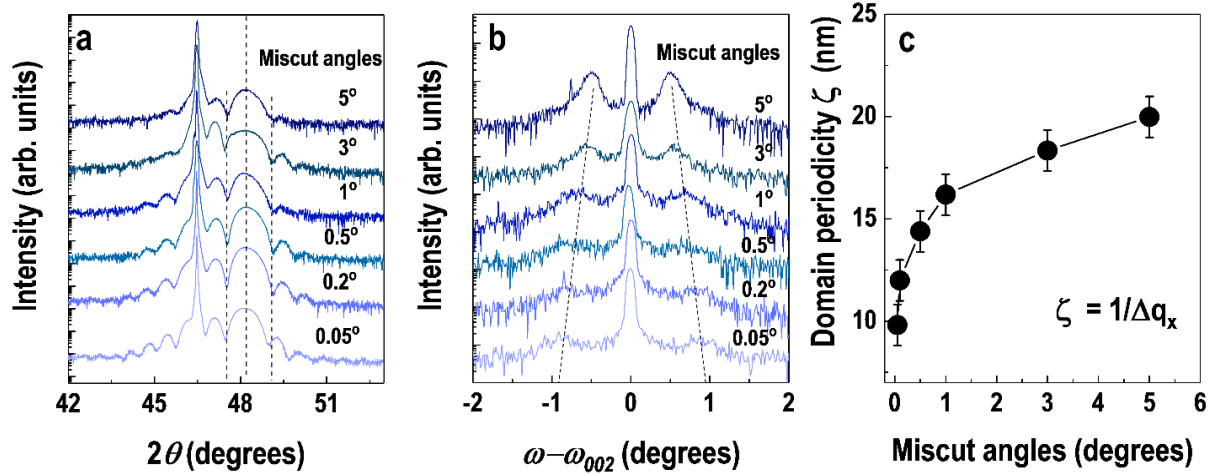

**Fig. S3 | Evolution of the 1D twinning domain in LCO films grown on STO substrates with different miscut angles.** The miscut angle is defined as the angle between the surface's normal and crystallographic [001] orientation. All miscut directions are in parallel to the in-plane [100] direction. The miscut angle of STO substrates is further confirmed by measuring the rocking curves of the STO 002 reflection. The topography of vicinal STO substrates with miscut angles from 0.05º to 5º was checked by AFM. The averaged width of STO terraces reduces significantly as the miscut angle increases. **a**. XRD $\theta$–$2\theta$ scans of the LCO films grown on vicinal STO substrates indicate all LCO films exhibit high crystalline quality and under the same strain state independent of the miscut angle. **b**. Rocking curves of the 002 reflections from the LCO films grown on STO substrates with different miscut angles. **c**. The twinning domain periodicity ($\zeta$) increases as the miscut angle increases. For the vicinal STO substrates, the fourfold symmetry at the surface plane is broken by exposing the facets at the step and terraces. The nuclei sites will initially form at the terraces and then the step-flow growth of LCO films is along the direction perpendicular to the steps. The larger miscut angle of STO substrate is, the smaller terrace width, which in turn gives rise to the shorter ion migration distance on the surface. Therefore, it is easier to form wider 1D twinning domains on the STO substrates with larger miscut angle.

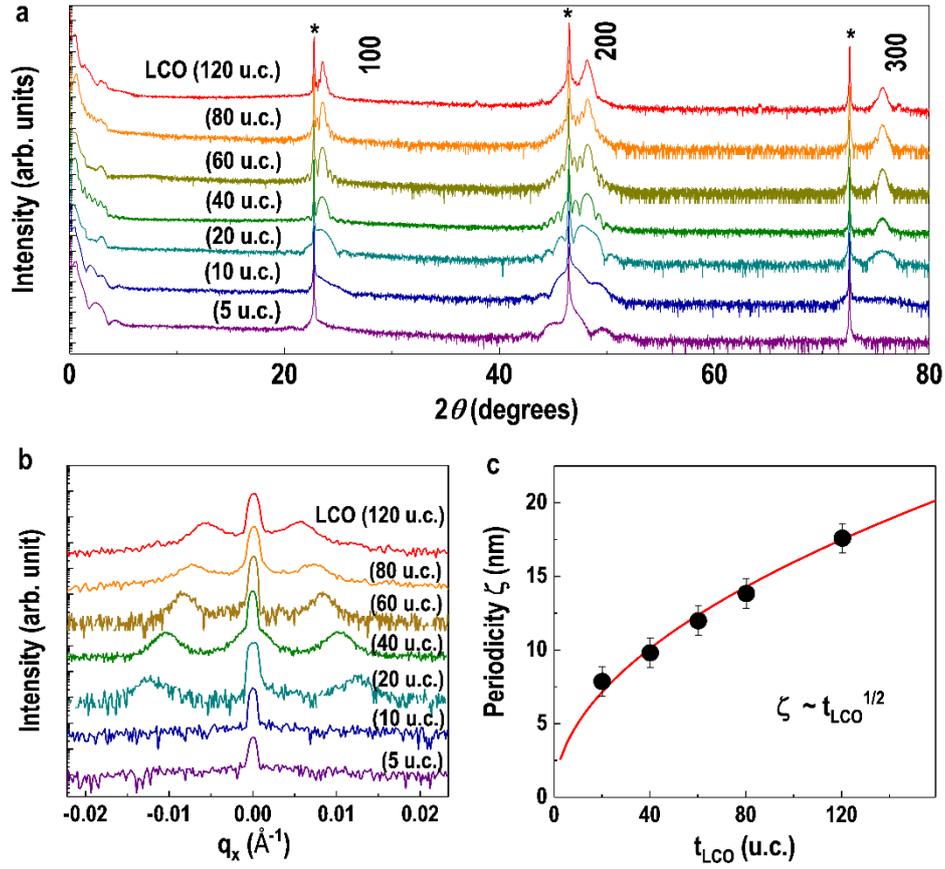

**Fig. S4 | Thickness dependent twinning domain periodicity in LCO films.** LCO films with thickness from 1.9 to 45.2 nm (~ 5 to 120 u.c.) are grown on the STO substrates. **a**. XRD $\theta$–$2\theta$ scans of LCO films indicate all films are epitaxial grown and of high crystalline quality. **b**. Rocking curves around 002 reflections of the LCO films with various thickness. For LCO thin films with thickness below 10 u.c., no splitting satellite peak is observed, suggesting no sizable twinning domain forms below this film thickness. The LCO ultrathin films suffer a large shear strain and have pseudotetragonal lattice structure due to the strong clamping with the cubic STO substrate / capping layer (*17*). The structural transition between the interfacial pseudotetragonal phase and monoclinic phase is gradual. The shear strain starts to relax when the LCO film thickness increases above 10 u.c., leading to the formation of ferroelastic twinning domains. **c**. The twinning domain periodicity ($\zeta$) increases as the square root of the film thickness as described by $\zeta \sim t_{LCO}^{1/2}$. This behavior agrees well with the thermodynamic consideration for the domain formation in epitaxial thin films. Similar evolution of twinning domain periodicity with film thickness has been observed for the $BiFeO_3$, $La_{1-x}Sr_xMnO_3$, and $SrRuO_3$ films previously.

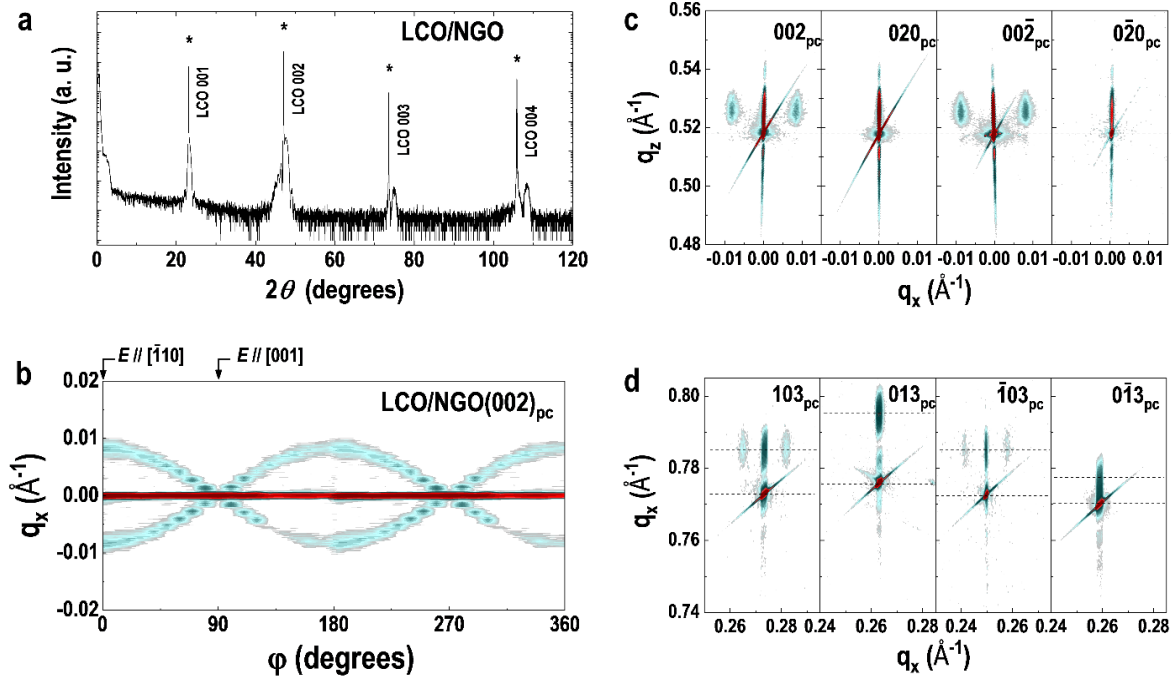

**Fig. S5 | 1D twinning domain in LCO films grown on NGO substrates.** NGO has an orthorhombic crystal structure with lattice constants $a$ = 5.431, $b$ = 5.499, and $c$ = 7.708 Å. It can be described as a pseudocubic lattice with a lattice constant of 3.861 Å. A LCO film was epitaxially grown on a (110)-oriented NGO substrate, as shown in **a**. The in-plane lattice parameter along the [001] and [$\bar{1}$10] orientation is different. Besides that, the NGO possesses an $c^+a^-a^-$ rotation pattern (Glazer notation) with a larger Ga-O-Ga bond angle along the [001] direction than that along the [$\bar{1}$10] direction (*34*). The anisotropic in-plane lattice mismatch and together with the octahedral tilt pattern will cause a unidirectional structural distortion of the LCO film along the [$\bar{1}$10] direction. This change of atomic structure is a result of the anisotropic shear strain from the dissimilar lattice structures. The in-plane anisotropic strain field at the LCO/NGO interface is similar to that of the LCO film grown on the well-defined terraced STO substrates. The former case is defined naturally by its crystal structure (orthorhombic structure), while the latter case is artificially created anisotropic strain by the long continuous step-and-terrace surface morphology. **b**. Rocking curve scans around the 002 reflection of the LCO films as a function of in-plane rotation angle ($\varphi$). We use the reciprocal space wavevector to show the position of splitting satellite peaks, from which the twinning domain period can be calculated. The $\Delta q_x$ ~ 0.006 Å$^{-1}$, corresponding to a twinning domain periodicity ($\zeta$) ~ 16.6 nm. Similar to the LCO/STO case (**Figure 1c**), a cosine modulation of the satellite peak position was observed, indicating the arrangement of these domain twinnings are 1D and perpendicular to the [$\bar{1}$10] orientation of NGO. **c – d**. RSMs from 0*kl* and *hk*3 reflections of a LCO film by successive rotating the sample at a step of 90° with

respect to the surface's normal, respectively. Only 00$l$ and $h$03 diffraction patterns show the satellite peaks, while the others are clearly absent. Therefore, the 1D twinning domain structure can be imposed by the selection of orthorhombic substrate.

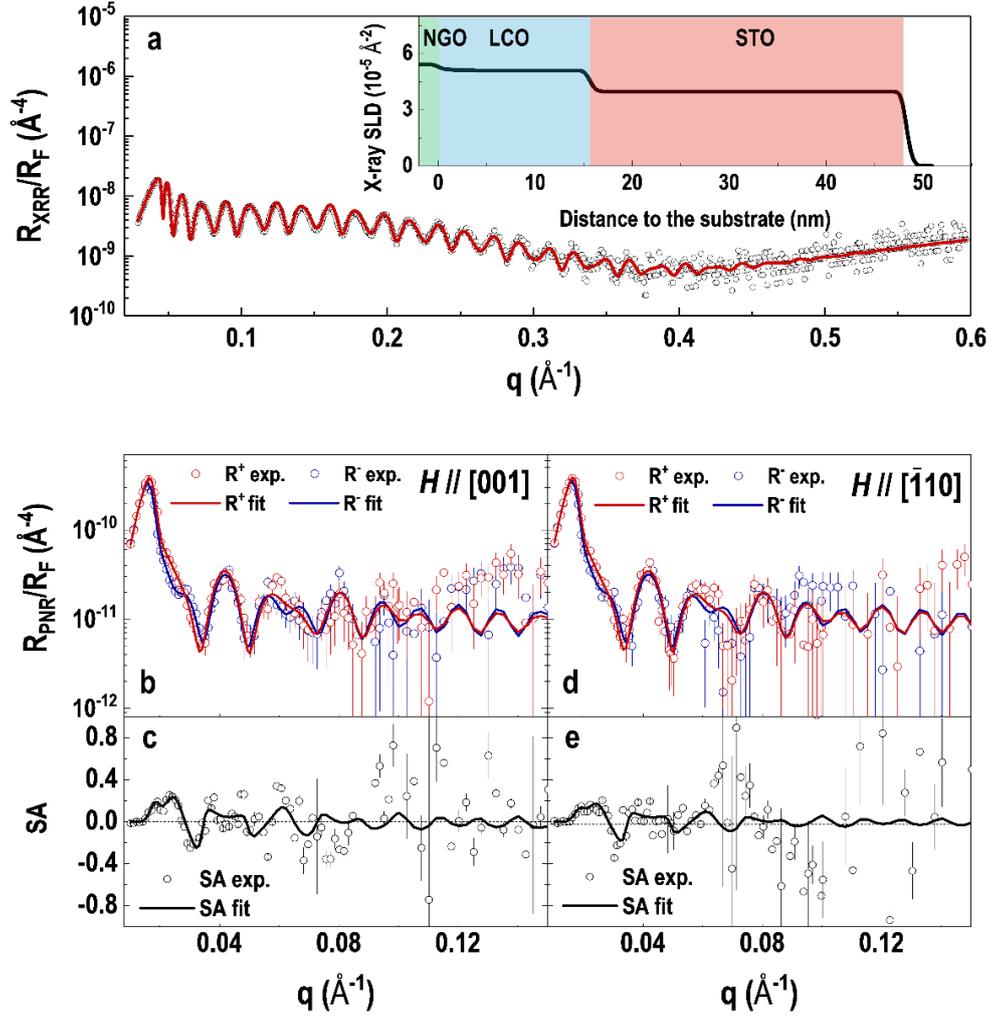

**Fig. S6 | In-plane magnetic anisotropy in LCO films grown on NGO substrates.** Due to the strong paramagnetic background from the NGO substrate and the large coercive field of LCO films, it is challenging to measure the magnetic properties of an LCO film grown on a NGO using SQUID magnetometry. We performed polarized neutron reflectivity (PNR) measurements on a STO/LCO bilayer sample grown on NGO to quantitatively determine the in-plane magnetization of LCO film. Prior to the PNR measurements, the chemical depth profile (inset of **a**) was obtained from fitting a model to the XRR data, as shown in **a**. The open circular symbol represents the experiment data and the solid line is the best fit. This model constrained the chemical depth profile used for fitting the neutron reflectivity. The PNR measurements were conducted at 10 K under a magnetic field $H = 3$ T. The sample was field-cooled when the magnetic field applied along the [001] and [$\bar{1}$10] orientations, respectively. The specular neutron reflectivity (**b** and **d**) is plotted as a function of the wave vector transfer $q$ ($=4\pi sin\theta_i/\lambda$) and normalized to the asymptotic value of the Fresnel reflectivity $R_F = (16\pi^2/q^4)$ for the spin-up ($R^+$) and spin-down ($R^-$) polarized neutrons, where

$\theta_i$ is the incident angle and $\lambda$ is the neutron wavelength. Solid lines are the best fit to the PNR data. The spin asymmetry (SA) and their corresponding fit are summarized (shown in **c** and **e**). The error bars in the **b** – **e** represent one standard deviation. The fitting results yield to $\chi^2$ metrics of 2.31 and 1.88 for the magnetic fields in parallel to the [001] and [$\bar{1}$10] orientations, respectively. The depth profiles of nuclear and magnetic scattering length density for the STO/LCO bilayer are plotted in **Figure 4c** and **4d** of main text. The in-plane magnetization of LCO film-bulk region along the [001] and [$\bar{1}$10] orientations are (0.70 ± 0.05) and (0.54 ± 0.02) $\mu_B$/Co at $H$ = 3 T, respectively. The PNR measurements were repeated on a second STO/LCO sample under the applied magnetic field of 0.7 T (not shown). The in-plane magnetization of the LCO film along the [001] and [$\bar{1}$10] orientations are (0.55 ± 0.05) and (0.23 ± 0.03) $\mu_B$/Co at $H$ = 0.7 T, respectively. The PNR results clearly show a sizable in-plane magnetic anisotropy in the LCO films grown on NGO substrates, further strengthening the strong modulation of in-plane magnetization by the unidirectional structural distortion in the LCO films.

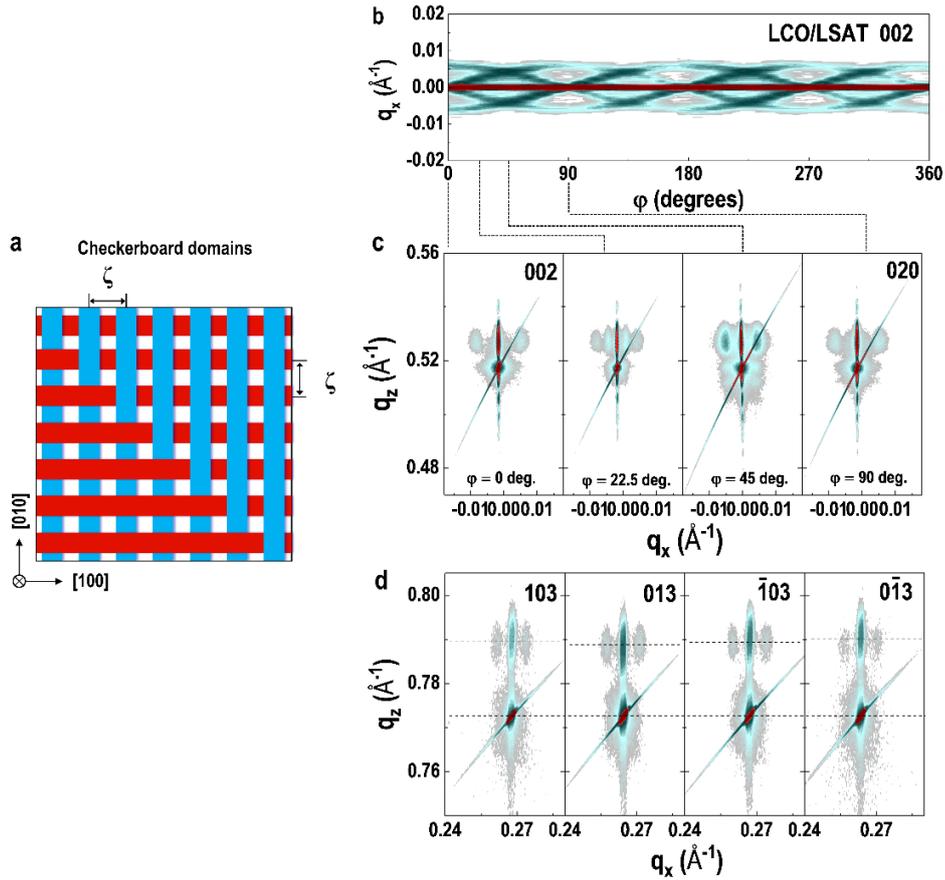

**Fig. S7 | Checkerboard-like twinning domains observed in LCO films on LSAT substrates.** The LSAT single crystal has a cubic lattice structure. When it is used as a substrate, it is normally not pre-treated prior to the thin film deposition due to cation segregations leading pure surface quality, thus no regular step-and-terrace feature on the surface was observed (checked by AFM). The formation of twinning domains along the [100] and [010] directions has the equal probability. Therefore, the checkerboard-like domains can be stabilized on the LSAT. **a**. Schematic of the checkerboard-like domains with averaged periodicity ($\zeta$). The structural modulation of a LCO film is bidirectional and overlapping each other. **b**. Rocking curves around the LCO 002 reflections as a function of in-plane rotation angle ($\varphi$). The real space angles are transformed into the reciprocal space wavevectors. The domain periodicity ($\zeta$) is calculated to be 12.1 nm ($\Delta q_x \sim 0.0083$ Å$^{-1}$). Both sine and cosine modulations of the satellite peak positions are observed, evidencing two kinds of twinning domains coexist in LCO films on LSAT substrates. These domains rotate 90° with respect to the other and are of approximately equal population (with the same reflection intensity). **c**. RSMs around the substrate's 0$kl$ reflection through successive rotating the sample by 0°, 22.5°, 45°, and 90° with respect to the surface's normal. The strongest reflection from the splitting satellite peaks was observed when the X-ray beam was aligned along the [110] diagonal direction (45°) because both twinning domains contribute their reflections at the

same reciprocal spacing. **d**. RSMs of the LCO film around the substrate's *hk*3 reflection. The splitting satellite peaks with the same intensity and reciprocal spacing are observed in all RSMs, again, indicating the equal distribution of twinning domains along the [100] and [010] orientations.

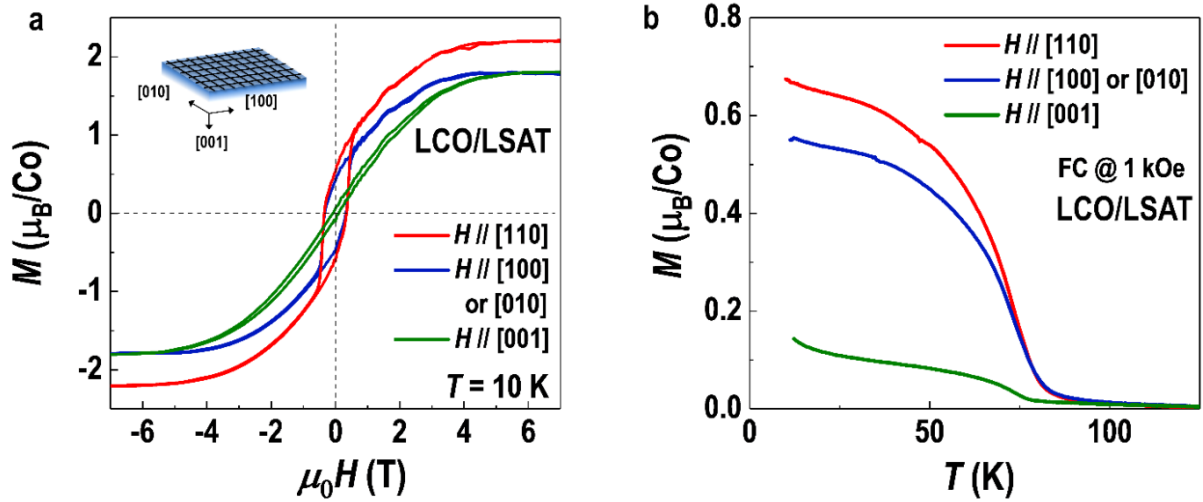

**Fig. S8 | Magnetic properties of LCO films on LSAT substrates with checkerboard-like twinning domains**. **a.** $M(H)$ hysteresis loops and **b.** $M(T)$ curves measured from LCO films on LSAT for $H$ is applied along the [010] or [100], [110], and [001] orientations. All magnetic hysteresis loops are recorded at 10 K with magnetic fields of ± 7 T. For the $M(T)$ curves, the cooling field was set at $H = 0.1$ T and all measurements were taken while warming. The nonlinear diamagnetic background from the substrate was subtracted by measuring the magnetization from a bare LSAT substrate separately. The magnetic hard axis of LCO films on LSAT is along the out-of-plane direction, similar to the LCO films on STO. The LCO film shows strong in-plane anisotropy. When the $H$ is applied parallel to one of the twinning domains' orientation, for instance the [100] or [010] direction, the magnetization is comparably small; however, when the $H$ is applied along the [110] orientation, *i. e.*, 45º with respect to both twinning domains, the magnetic response of LCO film reaches its maximum value. The magnetic responses of the LCO film along the [100] and [010] orientations are nearly identical (shown in blue curve in **a** and **b**). The in-plane magnetization is strongly correlated with the orientation of twinning domains.

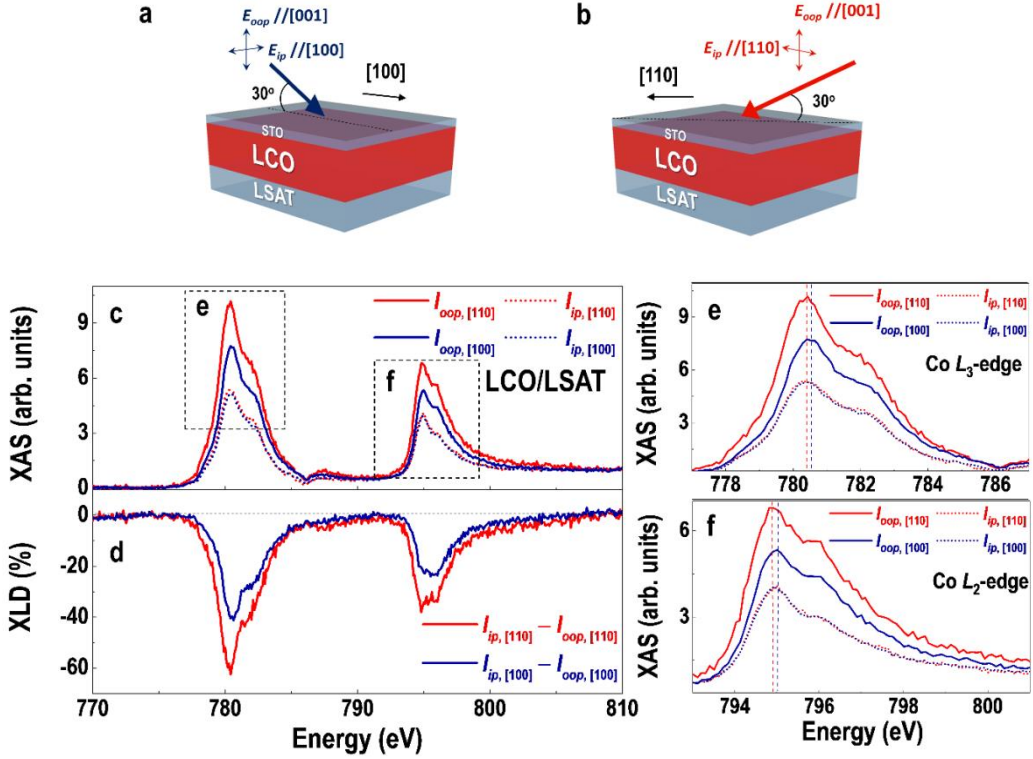

**Fig. S9 | Anisotropic electronic states in LCO films on LSAT substrates. a** and **b.** Schematics of the scattering geometry for XAS and XLD measurements when the X-ray beam is in parallel to the [100] and [110] directions, respectively. **c**. XAS of LCO film for Co $L$-edges measured by the out-of-plane ($I_{oop}$, solid lines, the linear polarization vector $E_{oop}$//[001]) and in-plane ($I_{ip}$, dashed lines, $E_{ip}$// [100] or [110]) linearly polarized photons. The lower energy of the $I_{ip}$ absorption spectra reveals that the electrons will preferentially occupy the $d_{x^2-y^2}$ orbital rather than the $d_{3z^2-r^2}$ orbital. XAS intensities of the $I_{oop}$ along the (100) and (110) planes are dramatically different. The black dashed rectangles in **c** highlight the Co $L_3$-edge (**e**) and Co $L_2$-edge (**f**). Direct inspection of the energy positions for the $I_{oop,\,[110]}$ and $I_{oop,\,[100]}$ are ~ 780.53 eV and ~ 780.42 eV, respectively. This result reveals the X-ray absorption in the (100) plane is ~ 0.11 eV lower in energy than that in the (110) plane. The difference implies a higher electron occupancy of the $d_{3z^2-r^2}$ orbital in the (100) plane than that in the (110) plane. From the peak energy shift of XAS, we could estimate the $e_g$ band splitting ($\Delta e_g$) between the $d_{3z^2-r^2}$ and $d_{x^2-y^2}$ orbitals are anisotropic in the different X-ray scattering planes. **d**. A comparison of XLD from the LCO films in the (100) and (110) planes. We find the XLD in the (100) plane is ~ 20% lower than that in the (110) plane. These observations suggest both anisotropic band splitting and orbital polarization arise from the anisotropic structural modulation induced by the formation of twinning domains in a LCO film.

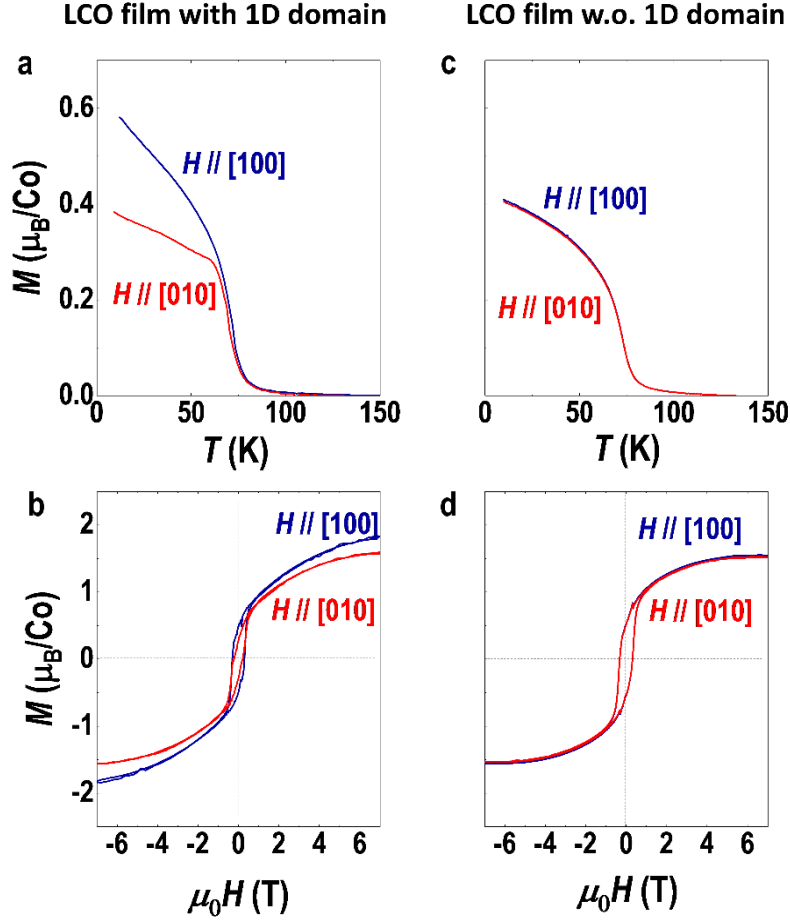

**Fig. S10 | Comparison of the in-plane magnetic anisotropy in LCO films with and without (w.o.) 1D twinning domains. a.** $M(T)$ curves and **b.** $M(H)$ hysteresis loops of LCO films with 1D twinning domains (**Figure 3**). We find a clear in-plane magnetic anisotropy in the LCO film. The magnetic response of the LCO film along the [010] orientation is smaller than that along the [100] orientation, revealing a strong modulation of magnetization by the unidirectional structural distortion. To further confirm the correlation between the twinning domains and magnetic anisotropy, we selected a pre-treated STO substrate with randomly oriented step-and-terrace feature on the surface (this is unlike the LSAT substrates). The LCO films grown on this STO substrate will not possess well-defined 1D or checkerboard-like twinning domains (confirmed by XRD measurements). **c.** $M(T)$ curves and **d.** $M(H)$ hysteresis loops of this LCO film show identical magnetization along [100] and [010] orientations. The saturation magnetization and the line shape of the $M(T)$ and $M(H)$ curves are consistent with previous reports (*13, 14, 17*). Therefore, our results clearly demonstrate another structural degree of freedom, e.g., the unidirectional lattice distortion, to effectively tune the anisotropic in-plane magnetization of the LCO films.